\documentclass[10pt,conference]{IEEEtran}
\usepackage{amsmath,amsfonts,algorithmic,algorithm,array,textcomp,stfloats,url,verbatim,graphicx,cite,amssymb,amsthm,algorithmic,graphicx,textcomp,xcolor,float,tabularx,epstopdf,graphicx,colortbl,blkarray,arydshln,mathtools,nicematrix,physics,algorithm,bm,authblk}
\usepackage[short,c2]{optidef}
\usepackage{tikz}

\DeclarePairedDelimiter\ceil{\lceil}{\rceil}
\DeclarePairedDelimiter\floor{\lfloor}{\rfloor}

\IEEEoverridecommandlockouts
\allowdisplaybreaks

\begin{document}

\title{Scalable Cyclic Schedulers for Age of Information Optimization in Large-Scale Status Update Systems}

\author[1]{Nail Akar}
\author[2]{Sahan Liyanaarachchi}
\author[2]{Sennur Ulukus}

\affil[1]{\normalsize Bilkent University, Ankara, T\"{u}rkiye}
\affil[2]{\normalsize University of Maryland, College Park, MD, USA}

\maketitle

\let\thefootnote\relax\footnotetext{This work is done when N.~Akar is on sabbatical leave as a visiting professor at University of Maryland, MD, USA, which is supported in part by the Scientific and Technological Research Council of T\"{u}rkiye  (T\"{u}bitak) 2219-International Postdoctoral Research Fellowship Program.}

\begin{abstract}
We study cyclic scheduling for generate-at-will (GAW) multi-source status update systems with heterogeneous service times and packet drop probabilities, with the aim of minimizing the weighted sum age of information (AoI), called system AoI, or the weighted sum peak AoI (PAoI), called system PAoI. In particular, we obtain well-performing cyclic schedulers which can easily scale to thousands of information sources and which also have low online implementation complexity. The proposed schedulers are comparatively studied against existing scheduling algorithms in terms of computational load and system AoI/PAoI performance, to validate their effectiveness.
\end{abstract}

\section{Introduction}
Research on timely status updates in networked control and remote monitoring systems has recently gained significant momentum. A multi-source status update system consists of a number of information sources, a server, and a destination(s). Each information source samples an associated random process with the sample values written into information packets. These packets are then collected by a server which is tasked with forwarding these packets to the destination \cite{RoyYates__AgeOfInfo_Survey}. In GAW systems, the server decides which information source to collect a fresh packet from, at a scheduling instant. Therefore, scheduling pertains to the collection phase for GAW systems. 

The age of information (AoI) process has recently been proposed to quantify information freshness in status update systems \cite{kaul_etal_infocom12}. In particular, the AoI process for an information source-$n$, $n \in \{1,2,\ldots,N\}$, denoted by $\Delta_n(t),t \geq 0$, is defined as $\Delta_n(t)= t-g_n(t)$ where $g_n(t)$ refers to the generation time of the last status update packet received at the destination from the same source.  The continuous-time continuous-valued AoI process $\Delta_n(t)$ increases with unit slope except for packet reception instances at which the process experiences an abrupt drop to a value which is the system time of the received packet. An alternative process is the peak AoI (PAoI) process $\Phi_n(k), k \in \mathbb{Z}^+$, which is obtained by sampling the AoI process of source-$n$ at the embedded epoch of the $k$th packet reception instant \cite{costa_peak}. The performance metrics used are the weighted sum of the mean values of the AoI and the PAoI processes, which are called the system AoI and the system PAoI, respectively, and the goal of the current paper is to develop schedulers that can scale to thousands of sources while attempting to minimize these metrics.

    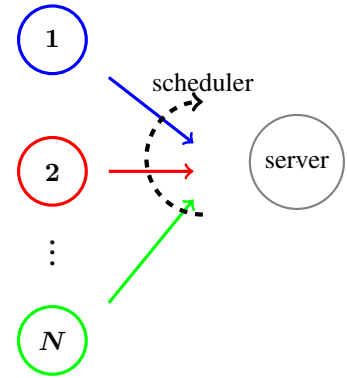
\begin{figure}[t]
    \centering
    \begin{tikzpicture}[scale=0.25]
    \tikzstyle{note} = [rectangle, dashed, draw, fill=white, font=\footnotesize,
    text width=20em, text centered, rounded corners, minimum height=18em]
    \draw[very thick, red](4,2) circle (1.8);
    \draw[very thick, green](4,-7) circle (1.8);
    \draw (4,1) node[anchor=south] {{${\bm 2}$}} ;
    \draw[very thick,blue] (4,9) circle (1.8)  ;
    \draw (4,8.1) node[anchor=south] {{${\bm 1}$}} ;
    \draw (4,-8) node[anchor=south] {{${\bm N}$}} ;
    \draw (4,-3.5) node[anchor=south] {{\Large $\vdots$}};
    \draw[very thick,red,->] (7,2) -- (11.5,2) ;
    \draw[very thick,blue,->] (7,7) -- (11.5,3.5) ;
    \draw[very thick,green,->] (7,-5) -- (11.5,0.5) ;
    \draw[ultra thick, black, ->, dashed] (12,-0.3) arc (270:90:3) node[anchor=south] {{scheduler}};
    \draw[thick, gray](17,2.5) circle (2.5);
    \filldraw (17,2.5) circle (0.01) node[anchor=center] {{server}};	
    \end{tikzpicture}
    \caption{Information packets from $N$ information sources are collected by a server, e.g., base station (BS) of a cellular wireless network.}
    \vspace*{-0.5cm}
    \label{fig:system}
    \end{figure}	

In this paper, we study the scheduling problem in Fig.~\ref{fig:system} for information packets collected from $N$ heterogeneous information sources towards a server which is co-located with the monitor. The packet drop probability (resp. success probability) for source-$n$, $1 \leq n \leq N$ is denoted by $p_n$ where $0 \leq p_n < 1$  (resp. $u_n=1-p_n$). The case of fixed and homogeneous service times for all sources in this setting has been extensively studied in the literature \cite{RoyYates__AgeOfInfo_Survey, AoI_min__multiSource_DT_GAW}. On the other hand, in the current paper, we envision generally distributed and heterogeneous packet transmission times. Moreover, information on packet drops is not available at the scheduler which needs to make open-loop scheduling decisions which can be probabilistic or cyclic, depending on whether the scheduler chooses to serve information sources with certain probabilities \cite{akar_gamgam_comlet23}, or according to a fixed repeating pattern of transmissions \cite{eywa}. The online implementation complexity of probabilistic schedulers is $O(N)$ whereas cyclic scheduling requires only $O(1)$ operations for online implementation due to the use of fixed patterns. 

We focus on cyclic scheduling for system AoI/PAoI minimization since we are interested in building (offline) scalable high-performance schedules that can scale to thousands of sources. The alternative closed-loop schedulers use the instantaneous values of the AoI processes and positive/negative acknowledgements. Age-aware schedulers such as max-weight and Whittle-index policies are in this category \cite{Kadota__BWN_SchedulingPoliciesForMinimizing, maatouk_etal_TWC20,kriouile_etal_T-IT22}. However, unawareness of packet drops at the server side renders the use of closed-loop scheduling infeasible in our system model.

Our scheduling work is most closely related to \cite{eywa} which also obtains cyclic schedules  based on a discrete-time setting with deterministic service times whereas in our work, we study generally distributed heterogeneous service times which covers the system model of \cite{eywa} as a sub-case. Our work is also related to \cite{AKAR2023109668} and \cite{gamgam_etal_arxiv23}, which focus on the minimization of system PAoI and AoI, respectively, for GAW status update systems, but no packet drops are assumed in these works. The solution we obtain for system PAoI minimization is in the form of a
square root law, i.e., the scheduling frequency of a source is proportional to the square root of the product of three terms, namely, the weight, mean service time, and the reciprocal of the success probability of a given source. Similar square root laws have also been obtained in several other AoI settings; see for example \cite{yates_etal_isit17, melih2020infocom}. However, system AoI minimization problem does not lead to a similar square root law in our system model. 

Our contributions are as follows. 1) We present an analytical method to obtain the per-source mean PAoI/AoI values in a GAW status update system with heterogeneous packet service times and packet drop rates, which is novel to the best of our knowledge. 2) We propose cyclic schedulers for optimized system PAoI and AoI performance, which are based on convex optimization and well-established packet spreading algorithms.

\section{System Model} \label{sec:systemmodel}
We consider the status update system in Fig.~\ref{fig:system} with $N$ information sources indexed by $n=1,2,\ldots,N$. In the GAW model, the server schedules a source to generate and transmit its information packet towards the server where packet generation involves the sampling of an associated random process and placement of the sample value in an information packet. Service times of source-$n$ packets (time required for the transmission of source-$n$ information packets) to the server whether successful or unsuccessful, denoted by $S_n$, are assumed to have a general distribution with moment generating function (MGF) $G_n(s)=\mathbb{E}[e^{s S_n}]$, mean $s_n =\mathbb{E} [S_n] = G_n^{\prime}(0)$, second moment $q_n = \mathbb{E} [S_n^2]=G_n^{\prime\prime}(0)$, variance $v_n = q_n - s_n^2$, and squared coefficient of variation (scov) $c_n = \frac{v_n}{s_n^2}$. In fact, there is no need to know the MGF $G_n(s)$ but only the first two moments in the proposed method; the MGFs are only used in the proofs. Once the transmission is over, a packet is unsuccessful (successful) with probability $p_n < 1$ ($u_n=1-p_n$). 
However, the server is assumed to have no information on whether the transmitted packet was successful or not, due to its open-loop nature.

Let $S_{n,k}$ denote the service time of the $k$th successful transmission of source-$n$ and let $\tilde{S}_{n,k}$ denote the time duration between the end of the $k$th successful transmission and the beginning of the $(k+1)$st successful transmission of source-$n$. Note that $\tilde{S}_{n,k}$ is the sum of the service times of all information packets generated from sources other than source-$n$ between two consecutive successful transmissions of source-$n$, denoted by $\tilde{S}^o_{n,k}$, and also of the service times of unsuccessful packets belonging to source-$n$ itself, denoted by $\tilde{S}^u_{n,k}$. Clearly, $\tilde{S}_{n,k} = \tilde{S}^o_{n,k} + \tilde{S}^u_{n,k}$.  The $k$th cycle of the AoI process $\Delta_n(t)$ starts from the value $S_{n,k}$ and increases with unit slope for a duration of $\tilde{S}_{n,k} + S_{n,k+1}$ until it reaches the value $S_{n,k} + \tilde{S}_{n,k} + S_{n,k+1}$ at which point a new cycle is initiated which will start from the value $S_{n,k+1}$. On the other hand, the discrete-time continuous-valued PAoI process $\Phi_n(k)$ is equal to the peak value in cycle-$k$, i.e., $\Phi_n(k) =S_{n,k} + \tilde{S}_{n,k} + S_{n,k+1}$. Let $\Delta_n$ (resp. $\Phi_n$) denote the steady-state random variable for the associated AoI (resp. system PAoI) process observed at the server for source-$n$, $1 \leq n \leq N$. We define the system AoI (resp. system PAoI) as the mean of the random variable $\Delta = \sum_{n=1}^N w_n \Delta_n$ (resp. $\Phi = \sum_{n=1}^N w_n \Phi_n$), i.e.,
\begin{align}
 \mathbb{E} [\Delta]  =  \sum_{n=1}^N w_n \mathbb{E} [\Delta_n], \qquad 
 \mathbb{E} [\Phi] =  \sum_{n=1}^N w_n \mathbb{E} [\Phi_n],\label{eqn:systemPAoI}
\end{align}
where the normalized source weights $w_n$, $\sum_{n=1}^N w_n=1$, are used to prioritize the sources according to their urgency. 

Let the random variable $\tilde{S}_n$ (steady-state random variable associated with the random process $\tilde{S}_{n,k}$ as $k \rightarrow \infty$) have mean $\tilde{s}_n =\mathbb{E} [\tilde{S}_n]$, second moment $\tilde{q}_n = \mathbb{E} [\tilde{S}_n^2]$, variance $\tilde{v}_n = \tilde{q}_n - \tilde{s}_n^2$ and scov $\tilde{c}_n = \frac{\tilde{v}_n}{\tilde{s}_n^2}$. Based on \cite{gamgam_etal_arxiv23}, the mean AoI for source-$n$ can be written in terms of the first two moments of the random variables $S_n$ and $\tilde{S}_n$ as,
\begin{align}
  \mathbb{E}[\Delta_n]  & = \frac{2 s_n^2 + 4 s_n \tilde{s}_n + q_n + \tilde{q}_n}{2(s_n + \tilde{s}_n)} \label{exp_2mom} \\
  & = \frac{s_n^2 (c_n + 3) +\tilde{s}_n^2(\tilde{c}_n + 1) + 4 s_n \tilde{s}_n }{2(s_n + \tilde{s}_n)}. \label{exp_scov}
\end{align}
On the other hand, the mean PAoI for source-$n$ is written in terms of the means of the random variables $S_n$ and $\tilde{S}_n$ as,
\begin{align}
  \mathbb{E}[\Phi_n]  & = 2s_n + \tilde{s}_n. \label{exp_1mom}
\end{align}

We note two observations from the expressions \eqref{exp_2mom}-\eqref{exp_1mom}. First, the expression for mean PAoI only depends on the first moments of $S_n$ and $\tilde{S}_n$, and is  much simpler than the expression for mean AoI. Second, for a given scheduling policy, finding the mean PAoI (resp. AoI) for any source-$n$ reduces to finding the first moment (resp. first two moments) of $\tilde{S}_n$ which depend on how the scheduler is taking its actions.

Next, we describe the cyclic GAW (C-GAW) scheduler which is characterized with a pattern $P = [P_0 ~ P_1 \cdots P_{K-1} ]$ of size $K$ such that $P_k \in \{1,2,\ldots,N \}$. In particular, the C-GAW scheduler initiates the sampling and transmission of source-$P_k$ information packet at scheduling instant $k+iK, \; i \in \mathbb{Z^+}$. As an example, let $N=3$ and $P = [1 ~ 2 ~ 3 ~ 2 ]$. In this case, the open-loop cyclic scheduler will transmit the following sequence of source packets: $1,2,3,2,1,2,3,2,1,\ldots$
A pattern is said to be feasible if each of the $N$ sources appears at least once in the pattern. On the other hand, the probabilistic GAW (P-GAW) scheduler chooses an information source for sampling and transmission with probability $r_n$. For both C-GAW and P-GAW, a new scheduling instant is initiated after the random service time $S_n$.

\section{Analytical Method} \label{sec:analysis}
We now present the analytical method which computes $\tilde{s}_n$ and $\tilde{q}_n$ for a given pattern $P$, which in turn gives the mean AoI and mean PAoI through \eqref{exp_2mom} and \eqref{exp_1mom}, respectively. For this purpose, let $\alpha_n$, $n=1,\ldots,N$ denote the number of times that source-$n$ appears in the pattern $P$. Note that $\alpha_n > 0$, for all $n$, since otherwise the weighted AoI/PAoI would be unbounded. Let $P_{n,k}$, $k=0,1,\ldots,\alpha_n-1,$ denote the sub-pattern obtained by deleting all entries in the original pattern $P$ except for the entries between the $k$th and $(k+1)$st $\left( \text{modulo } \alpha_n \right)$  appearances of source-$n$, i.e., the instances at which $P_k=n$ and $P_{k+1}=n$, respectively, excluding the end points. 
 Let $\alpha_{n,k,m}$ denote the number of times source-$m$ appears in the sub-pattern $P_{n,k}$. Let $s_{n,k}$ and $v_{n,k}$ denote the mean and variance, respectively, of the sum of the service times, denoted by $H_{n,k}$, of all the sources contained in sub-pattern $P_{n,k}$, which are obtained by summing the means and variances, respectively, of the service times of these sources, 
\begin{align}
  s_{n,k} & = \sum_{m \neq n} \alpha_{n,k,m} s_m, \qquad
  v_{n,k} = \sum_{m \neq n}  \alpha_{n,k,m} v_m.
\end{align}
Also, let $q_{n,k} = v_{n,k} + s_{n,k}^2$ and $G_{n,k}(s) $ denote the second moment and MGF, respectively, of $H_{n,k}$, i.e., 
\begin{align}
  G_{n,k}(s) & = \mathbb{E} [e^{sH_{n,k}}] = \prod\limits_{m \neq n} G_m(s)^ {\alpha_{n,k,m}}.
\end{align}
Let $\tilde{H}_{n,k}$ denote the time between two consecutive successful transmissions of source-$n$ (excluding the end points) given that the first successful transmission occurs at the $k$th appearance of source-$n$ in the pattern $P$ and let
$\tilde{G}_{n,k}(s) = \mathbb{E} [e^{s \tilde{H}_{n,k}}]$ denote its MGF. From the total law of expectation, we have,
\begin{align}
 & \!\frac{\tilde{G}_{n,k}(s)}{u_n} =  G_{n,k}(s) 
                    +  p_n G_n(s)G_{n,k}(s)  G_{n,k+1}(s) \nonumber\\
                      & \! + \! p_n^2 G_n(s)^2 G_{n,k}(s)  G_{n,k+1}(s) G_{n,k+2}(s) \! + \! \ldots \nonumber\\
                      & \! + \! p_n^{\alpha_n-1} G_n(s)^{\alpha_n-1} {\prod_{m=1}^{\alpha_n} G_{n,m-1}(s)} \nonumber\\
                      & \! + \! p_n^{\alpha_n} G_n(s)^{\alpha_n}  G_{n,k}(s) {\prod_{m=1}^{\alpha_n} G_{n,m-1}(s)} \nonumber\\
                      & \! + \! p_n^{\alpha_n} G_n(s)^{\alpha_n +1} G_{n,k}(s) G_{n,k+1}(s) {\prod_{m=1}^{\alpha_n} G_{n,m-1}(s)} \! + \! \ldots \label{mgf75}
\end{align}
where $G_{n,k}(s)=G_{n,k+\alpha_n}(s)$, $0 \leq k \leq \alpha_n$ by convention. A successful source-$n$ packet belongs to a transmission opportunity at appearance-$k$, $k=0,1,\ldots,\alpha_n$, uniformly. Therefore, the MGF $\tilde{G}_{n}(s)$ of the random variable $\tilde{S}_n$ is written as,
\begin{align}
 \tilde{G}_{n}(s) &= \frac{1}{\alpha_n} \sum_{k=1}^{\alpha_n} \tilde{G}_{n,k-1}(s). \label{nail76}           
\end{align}
By differentiating \eqref{mgf75} with respect to the variable $s$ and evaluating the result at $s=0$, we obtain,
\begin{align}
 \tilde{s}_n &=  \mathbb{E} [\tilde{S}_n] = \tilde{G}_n'(0) = \frac{1}{\alpha_n} \sum_{k=1}^{\alpha_n} \tilde{G}'_{n,k-1}(0), \\
 & = \frac{1}{\alpha_n} \sum_{k=1}^{\alpha_n} \left( \frac{p_n s_n}{u_n} + \frac{\sum_{j=1}^{\alpha_n} p_n^{j-1} s_{n,k+j-1}}{1-p_n^{\alpha_n}} \right),\\
 & =\frac{1}{u_n} \left( p_n s_n +  \frac{1}{\alpha_n} \sum_{k=1}^{\alpha_n} s_{n,k-1} \right).
 \label{nail77}           
\end{align}
Again, in the equations above, we used $s_{n,k}=s_{n,k+\alpha_n}$, $0 \leq k \leq \alpha_n-1$ by convention. 

Obtaining $\tilde{q}_n$ is more involved than obtaining $\tilde{s}_n$ for which an explicit expression was given in \eqref{nail77}. We now present a numerical method to obtain $\tilde{q}_n$ for which purpose we rearrange the terms in \eqref{mgf75} to write $\tilde{G}_{n,k}(s)$ as,
\begin{align}
 \tilde{G}_{n,k}(s) & = 
  \frac{u_n \sum_{j=1}^{\alpha_n} p_n^{j-1} G_n(s)^{j-1} \prod_{l=1}^j G_{n,k+l-1}(s) }{1 - p_n^{\alpha_n}G_n(s)^{\alpha_n} {\prod_{m=1}^{\alpha_n} G_{n,m-1}(s)}}.                  
\end{align}
Ignoring the higher order terms (since they will not have any effect on $\tilde{G}_{n,k}''(0)$), we write,
\begin{align}
  G_n(s) & = 1 + s_n s + \frac{q_n}{2} s^2 + O(s^3), \\
  G_{n,k}(s) & = 1 + s_{n,k} s + \frac{q_{n,k}}{2} s^2 + O(s^3),
\end{align}
from which we can write, 
\begin{align}
 \tilde{G}_{n,k}(s) & = \frac{(1-p_n^{\alpha_n}) + a_{n,k} s + {b_{n,k}} s^2 + O(s^3)}{(1-p_n^{\alpha_n}) + c_{n,k} s + d_{n,k} s^2 + O(s^3)}, 
\end{align}
where the coefficients $a_{n,k}$, $b_{n,k}$, $c_{n,k}$ and $d_{n,k}$ can be found numerically, e.g., by applying the convolution operator for the product of MGFs, and the $O(s^3)$ terms vanish when they are differentiated with respect to $s$ once or twice, and evaluated at $s=0$. Using the quotient rule on the derivative of the ratio of two differentiable functions, we have,
\begin{align}
 \tilde{q}_n &=  \mathbb{E} [\tilde{S}_n^2] = \tilde{G}_n''(0) = \frac{1}{\alpha_n} \sum_{k=1}^{\alpha_n} \tilde{G}''_{n,k-1}(0), \\
 & = \frac{1}{\alpha_n} \sum_{k=1}^{\alpha_n} \left( \frac{2c_{n,k}(c_{n,k}-a_{n,k})} {(1-p_n^{\alpha_n})^2} +
 \frac{2(b_{n,k}-d_{n,k})} {(1-p_n^{\alpha_n})} \right). \! 
 \label{nail79}           
\end{align}
Substituting the values of $\tilde{s}_n$ and $\tilde{q}_n$ obtained in \eqref{nail77} and \eqref{nail79}, respectively, in \eqref{exp_2mom} and \eqref{exp_1mom}, gives the mean AoI and mean PAoI for source-$n$, respectively, concluding the presentation of the analytical method. 

The computational complexity of finding the system AoI/PAoI of a pattern with size $K$ with $N$ sources is ${O}(NK)$.

\section{System PAoI and AoI Optimization} \label{sec:optimization}

\subsection{System PAoI Optimization}
Let $\tau_n$, $0 < \tau_n < 1$, $n \in \{1,\ldots,N\}$, denote the link utilization of source-$n$, i.e., long-term fraction of time
that the wireless link is occupied with the transmission of packets (successful or unsuccessful) from source-$n$. Since we focus on work-conserving servers, we have $\sum_{n=1}^N \tau_n =1$. Moreover, let the random variable $\tilde{S}_n^u$ (steady-state random variable associated with the random process $\tilde{S}_{n,k}^u$ as $k \rightarrow \infty$) have mean $\tilde{s}_n^u =\mathbb{E} [\tilde{S}_n^u]$. Then, it is not difficult to write,
\begin{align}
    \tau_n = \frac{s_n + s^u_n}{s_n + \tilde{s}_n}. \label{nail41}
\end{align}
However, it is easy to see that $s^u_n = \frac{p_n}{u_n} s_n$,
since for every successful source-$n$ transmission, we must have on the average $\frac{p_n}{u_n}$ unsuccessful transmissions from the same source. From \eqref{nail41}, we obtain $\tilde{s}_n$ in terms of $\tau_n$ as,
\begin{align}
  \tilde{s}_n & = \frac{s_n}{u_n \tau_n} - s_n, \label{nail43}  
\end{align}
which yields expressions for $\mathbb{E}[\Phi_n]$ and the system PAoI as,
\begin{align}
 \mathbb{E}[\Phi_n]  & = s_n + \frac{s_n}{u_n \tau_n}, \quad \mathbb{E}[\Phi]  =  \sum\limits_{n=1}^N \frac{w_n s_n}{u_n \tau_n} + \bar{s}, \label{nail44}  
\end{align}
where the fixed term $\bar{s} = \sum_{n=1}^N w_n s_n$ does not depend on $\tau_n$.

The minimization of the system PAoI in \eqref{nail44} subject to $\sum_{n=1}^N \tau_n =1$ is a convex optimization problem. To see this, the function $f(x)=\frac{1}{x}$ is a convex function of $x$ for $x>0$ and a non-negative weighted sum of convex functions is also convex \cite{boyd2004convex}. Therefore, we apply the Karush-Kuhn-Tucker (KKT) conditions \cite{boyd2004convex} on the system PAoI expression to  obtain the optimum utilization parameters $\tau_n^*$, $1 \leq n \leq N,$ that minimize the system PAoI,
\begin{align}
\tau_n^* & \propto \sqrt{\frac{w_n s_n}{u_n}}, \quad 1 \leq n \leq N. \label{eqn:optimum_utilization}
\end{align}
Recall that $f_n$ denotes the long-term frequency of packet transmissions from source-$n$. Since $\tau_n \propto f_n s_n$ (by definition), the optimum transmission frequency, denoted by $f_n^*$, $1 \leq n \leq N$, is given by the following closed-form expression,
\begin{align}
f_n^* & = {\sqrt{\frac{w_n}{s_n u_n}}}\left({\sum_{m=1}^N\sqrt{\frac{w_m}{s_m u_m}}}\right)^{-1}.
\label{eqn:optimum_frequency}
\end{align}

Note that a P-GAW scheduler that probabilistically schedules source-$n$ with probability $f_n^*$ is optimum in terms of weighted sum PAoI minimization. However, two observations stand out from this analysis. First, P-GAW scheduling requires $O(N)$ operations at a scheduling instant in contrast with C-GAW scheduling which requires $O(1)$ operations for online implementation. Moreover, it is known that probabilistic schedulers work poorly for AoI as shown in \cite{gamgam_etal_arxiv23}. Therefore, there is a need to develop C-GAW schedulers that attempt to attain the transmission frequency $f_n^*$ given in \eqref{eqn:optimum_frequency}.

As the next step for cyclic scheduling, we first need to write $f_n^* \approx K_n/K$ for integer $K_n$, $K$ for which an algorithm is proposed in Algorithm~\ref{alg:General} in terms of the algorithm parameter $\varepsilon \geq 0$. The case of $\varepsilon =0$ ensures that $K_n \geq 1, \; \forall n,$ and has a small pattern size $K$. When $\varepsilon$ increases, obviously the approximation $K_n/K$ of the optimum frequency $f_n^*$ improves but at the expense of increased $K$ and hence increased storage requirements. The computational complexity of Algorithm~\ref{alg:General} is ${O}( N\log N)$ due to the required sorting.

\begin{algorithm}[tbh]
    \caption{Pseudo-code for obtaining $K_n$ and $K$}
    \begin{algorithmic}
    \renewcommand{\algorithmicrequire}{\textbf{Input:}}
    \renewcommand{\algorithmicensure}{\textbf{Output:}}
        \REQUIRE $f_n^*, 1 \leq n \leq N$, $\varepsilon \geq 0$;
        \ENSURE $K_n,K$;
        \STATE {\bf Step 1:} Find $f_{min}^* =\min_n f_n^*$;
        \STATE {\bf Step 1:} Set
        $K = \frac{(1+\varepsilon)}{f_{min}}$;
        \STATE  {\bf Step 3:} Find $R = \sum_{n=1}^N \floor{K f_n}$;
        \STATE {\bf Step 4:} Sort all the sources in descending order according to the fractional part of $K f_n$ and set $K_n = \ceil{K f_n}$ for the top $K-R$ sources in this ordered list, otherwise set $K_n = \floor{K f_n}$.
    \end{algorithmic}
    \label{alg:General}
\end{algorithm}

What is now needed is a packet spreading algorithm which generates a transmission pattern in which source-$n$ appears $K_n$ times in it with pattern size $K=\sum_{n=1}^N K_n$, and all these appearances are as evenly spread as possible throughout the pattern. This problem with many variations had been studied in the context of \emph{internet scheduling} where the goal was to share the link bandwidth fairly among multiple flows carrying fixed-size or variable-size packets; see \cite{shreedhar_varghese_sigcomm95} and the references therein for a collection of research papers on fair link sharing.

The packet spreading algorithm we propose to use in this paper is based on the deficit round robin (DRR) algorithm proposed in \cite{shreedhar_varghese_sigcomm95} which has been successfully used in commercial routers due to its low computational complexity. DRR consists of rounds at each of which the deficit counters of each flow are incremented by the product of the so-called quantum and the weight of the flow. Subsequently, all the head-of-line packets waiting in the queue of each flow whose total packet size in bytes does not exceed the corresponding deficit counter, are served. The number of bytes that are served for a flow are then subtracted from the corresponding deficit counter, leaving a so-called deficit. In this way, multiple flows can be served in the same round. In our proposed spreading algorithm, our goal is to find a C-GAW scheduler that spreads out source-$n$ transmissions as much as possible while source-$n$ appears $K_n$ times in the pattern. Initially, all deficit counters, denoted by $B_n(t)$, $1 \leq n \leq N$, are set to zero. We modify the original DRR algorithm by allowing the value of the quantum change between rounds, which is slightly different than the original DRR scheduler \cite{shreedhar_varghese_sigcomm95}. Particularly, the quantum is chosen to ensure that one source is guaranteed to be inserted into the pattern at a given round while leaving zero deficit. The pseudo-code for the proposed packet spreading algorithm is described in Algorithm~\ref{alg:Spreading} with computational complexity ${O}(NK)$.

\begin{algorithm}[tb]
    \caption{Algorithm for constructing the pattern $P$}
    \begin{algorithmic}
    \renewcommand{\algorithmicrequire}{\textbf{Input:}}
    \renewcommand{\algorithmicensure}{\textbf{Output:}}
        \REQUIRE $N \geq 2$, $K_n, n=1,\ldots,N$, $K=\sum_{n=1}^N K_n$;
        \ENSURE Pattern $P$ of size $K$;
        \STATE $B_n(t) \gets 0$; $\;$ ($B_n(t)$: deficit counter for source-$n$)
        \FOR {round $k=0$ \TO round $K-1$}
        \STATE $m \gets \arg \min\limits_{ 1 \leq n \leq N} \frac{(1-B_n(t))K}{K_n}$;  (ties broken randomly)
        \STATE $Q \gets \frac{(1-B_m(t))K}{K_m}$; ($Q$: quantum)
        \FOR {$n=1$ \TO $N$}
        \STATE $B_n(t) = B_n(t) + Q \frac{K_n}{K}$; (update deficit counters)
        \ENDFOR
        \STATE $P(k) \gets m$; (insert source-$m$ in $P$)
        \STATE $B_m(t) \gets 0$;
        \ENDFOR
    \end{algorithmic}
    \label{alg:Spreading}
\end{algorithm}

To summarize the overall scheduling algorithm, termed as SPMS (System PAoI Minimizing Scheduler) in terms of the non-negative algorithm parameter $\varepsilon \in \mathbb{R} $, in the first step, the frequencies $f_n^*$ are obtained according to \eqref{eqn:optimum_frequency}. In the second step, the pattern size $K$ and the number of appearances of source-$n$ in the pattern, namely $K_n$, are obtained using Algorithm~\ref{alg:General} given $\varepsilon$ which plays out the trade-off between system PAoI performance and increased pattern storage requirements. In the final step, the pattern $P$ is constructed using the packet spreading algorithm of Algorithm~\ref{alg:Spreading}. This final step favors system AoI but does not have an impact on system PAoI.

\subsection{System AoI Optimization}
Next, we turn our attention to system AoI minimization in which case the source utilization variables $\tau_n$ affect directly the parameters $\tilde{s}_n$ but not the scov parameters $\tilde{c}_n$ in \eqref{exp_scov}. In fact, $\tilde{c}_n$s are complex functions of the scheduler itself and therefore it is difficult to pose a joint optimization problem involving both $\tilde{s}_n$ and $\tilde{c}_n$ for $1 \leq n \leq N$.
Note that for a fixed choice of $\tau_n$ values, the best choice for the minimization of system AoI would be to schedule the sources in such a way that $\tilde{c}_n$ parameters are as close to zero as possible, i.e., close to periodic transmissions of successful source-$n$ packets. However, this may not always be possible even in the case of deterministic service times. When source-$n$ transmissions are periodic, the time interval between two consecutive successful source-$n$ transmissions may not necessarily be deterministic, i.e., $\tilde{c}_n$ will be non-zero, which stems from non-zero packet error probability for source-$n$. Moreover, for source-$n$ transmissions to be periodic, the ratio $K/K_n$ needs to be an integer which may not necessarily hold in general. Even so, it may not be possible for source-$n$ transmissions to be periodic due to constraints from other sources. When the service times are random, this situation becomes  even more challenging.  

We first describe the optimization algorithm we formulate for given $\tilde{c}_n$. Substituting \eqref{nail43} into \eqref{exp_scov}, we obtain,
\begin{align}
2 w_n \mathbb{E}[\Delta_n]  = & \underbrace{w_n s_n u_n (c_n + \tilde{c}_n)}_{a_n} \tau_n  \nonumber \\
& + \underbrace{\frac{w_n s_n (1+\tilde{c}_n)}{u_n}}_{b_n} \frac{1}{\tau_n} + 2 w_n s_n (1 - \tilde{c}_n), \label{nail51}
\end{align}
which results in the following optimization problem for system AoI minimization,
\begin{mini}
    {\tau_n \geq 0}{\sum_{n=1}^{N} \left( a_n \tau_n +  b_n \frac{1}{\tau_n} \right) }
    {\label{Optimization1}}
    {}
    \addConstraint{ \sum_{n=1}^{N} \tau_n}{= 1},
\end{mini}
where the coefficients $a_n$ and $b_n$ are non-negative. The optimization problem \eqref{Optimization1} is convex and the KKT conditions result in the following condition, 
\begin{align}
    a_n - \frac{b_n}{\tau_n^2} & = a_m - \frac{b_m}{\tau_m^2} , \quad 1 \leq n,m \leq N,  
\end{align}
which together with the normalization constraint in \eqref{Optimization1} yields the following non-linear fixed point equation in the single unknown $x \in \mathbb{R}$,
\begin{align}
    f(x) & = \sum_{n=1}^N \sqrt{\frac{b_n}{a_n-x}} -1 = 0. \label{fixedpoint}
\end{align}
Let $a = \min\limits_{1 \leq n \leq N} a_n$. We note that the function $f(x) \in (-1,\infty)$ is a monotonically increasing function of $x \in (-\infty,a)$,
and therefore has a unique solution denoted by $x^*$ (can be found by bisection search) from which one can find the optimum coefficients $\tau_n^*$, $1\leq n \leq N$, from, 
\begin{align}
    \tau_n^* & = \sqrt{\frac{b_n}{a_n-x^*}}. 
\end{align}
Subsequently, the frequency parameters $f_n^*$ can be written as,
\begin{align}
    f_n^* & = {\frac{\tau_n^*}{s_n}} \left( {\sum\limits_{m=1}^N \frac{\tau_m^*}{s_m}} \right)^{-1}. \label{eqn:optimum_frequencyAoI}
\end{align}

Although, the procedure described above obtains the frequencies optimally, it does not lend itself to an algorithmic procedure to find the optimum pattern as in SPMS since $\tilde{c}_n$ are not known in advance. For this purpose, we propose the following fixed-point iterations to construct a transmission pattern for system AoI optimization. We first start with an initial value for $\tilde{c}_n^{(0)}=p_n$ which is based on the assumption that the scheduler is able to transmit source-$n$ packets with a period of $T_n = \frac{s_n}{\tau_n}$ which will ensure a link utilization of $\tau_n$. With this assumption in hand, $\tilde{S}_n$ behaves as $\tilde{S}_n = T_n X_n - S_n$ where $X_n$ is a geometrically distributed random variable with parameter $u_n$. Recalling that a geometrically distributed random variable with parameter $u_n$ has mean $\frac{1}{u_n}$ and variance $\frac{1-u_n}{u_n^2}$, we have,
\begin{align}
 \mathbb{E} [\tilde{S}_n]  = \frac{T_n}{u_n} - s_n, \qquad \text{Var} [\tilde{S}_n]  =  \frac{T_n^2 p_n}{u_n^2} + v_n. 
\end{align}
In a large-scale status update system, $T_n >> s_n$ and $T_n^2 >> v_n$. Hence, we propose to approximate $\tilde{c}_n = 
\frac{\text{Var} [\tilde{S}_n] }{ \mathbb{E} [\tilde{S}_n]^2}$ by $p_n$ which does not depend on $\tau_n$ which underlies the  initial choice of $\tilde{c}_n^{(0)}=p_n$. Subsequently, in iteration $\ell$, $1 \leq \ell\leq L$,
for a given value of $\tilde{c}_n^{(\ell-1)}$, we obtain the source-$n$ frequency $f_n^{*(\ell)}$ according to \eqref{eqn:optimum_frequencyAoI}. 
Then, for a given value of $\varepsilon \geq 0$ from a given set $\mathcal{E}$ of $\varepsilon$ values, we employ Algorithms~\ref{alg:General} and \ref{alg:Spreading} to obtain a pattern $P^{(\ell,\varepsilon)}$ in iteration $\ell$ whose corresponding system AoI $\mathbb{E}[\Delta^{(\ell,\varepsilon)}]$ and $\tilde{c}_n{^{(\ell,\varepsilon)}}$ values can be calculated based on the procedure described in Section~\ref{sec:analysis}. We repeat the process for each value $\varepsilon$ in the set $\mathcal{E}$ and choose the particular value of $\varepsilon$, denoted by $\varepsilon'$, resulting in a pattern $P^{(\ell)}=P^{(\ell,\varepsilon')}$ with the minimum system AoI in this round. Subsequently, the parameter $\tilde{c}_n{^{(\ell)}}=\tilde{c}_n{^{(\ell,\varepsilon')}}$ is fed as input to iteration $\ell+1$. This procedure repeats for $L$ iterations. The pattern $P^{(\ell)}$ which generates the lowest system AoI among all the iterations $1 \leq \ell \leq L$ is then output by our proposed algorithm, called SAMS (System AoI Minimizing Scheduler), which is a function of the algorithm parameters $\mathcal E$ and $L$. Hence, the proposed method SAMS should not be viewed as a vector fixed-point iteration but rather a search algorithm along the fixed-point iterations. For a given $\varepsilon$, the computational complexity of one SAMS iteration is ${O}(NK)$ due to the fact that Algorithm~\ref{alg:Spreading} dominates the overall execution time where $K$ is the size of the pattern produced by SAMS.

\section{Numerical Examples} \label{sec:Numerical}
In the first numerical example, we study a small-scale status update system containing $N=3$ sources with weights selected as $w_1=5w_2=25w_3$, deterministic service times with $s_1=5,s_2=2.5$, and no errors. The mean service time of source-3, $s_3$, is varied to assess the system AoI performance using three variations of SAMS and three benchmark policies. In SAMS-1, the algorithm parameters are selected as $\mathcal{E}=\{ 0 \},L=1$ whereas SAMS-2 employs a wider set ${\mathcal E} = \{ 0:0.2:2 \}$ for $\varepsilon$ but does not perform fixed-point iterations. Finally, SAMS-3 uses the same set ${\mathcal E}$ but employs fixed-point iterations with $L=3$. The benchmark policies are: (i) RR (round robin) policy. (ii) P-GAW$^*$ policy obtained by solving the system AoI for P-GAW using the technique of \cite{gamgam_etal_arxiv23} employing exhaustive search over the transmission probabilities $r_n$. (iii) IS (insertion search) iterative algorithm proposed in \cite{gamgam_etal_arxiv23} which starts with the RR policy initially which is equivalent to using the particular pattern $P = [1 ~ 2 ~ 3]$. At round $i+1, 3\leq i \leq I$, the IS method obtains a pattern of size $i+1$ from the pattern obtained in round $i$ by means of inserting the particular source transmission along with its position in the pattern resulting in the largest drop in the system AoI. The pattern with the lowest system AoI among the patterns of all rounds is then returned. It was shown in \cite{gamgam_etal_arxiv23} that IS produces the optimum pattern when $N=2$ when the IS search is terminated when system AoI cannot be improved at a given round.

\begin{figure}[t]
    \centering
     \vspace*{-0.5cm}
    \includegraphics[width=0.42\textwidth]{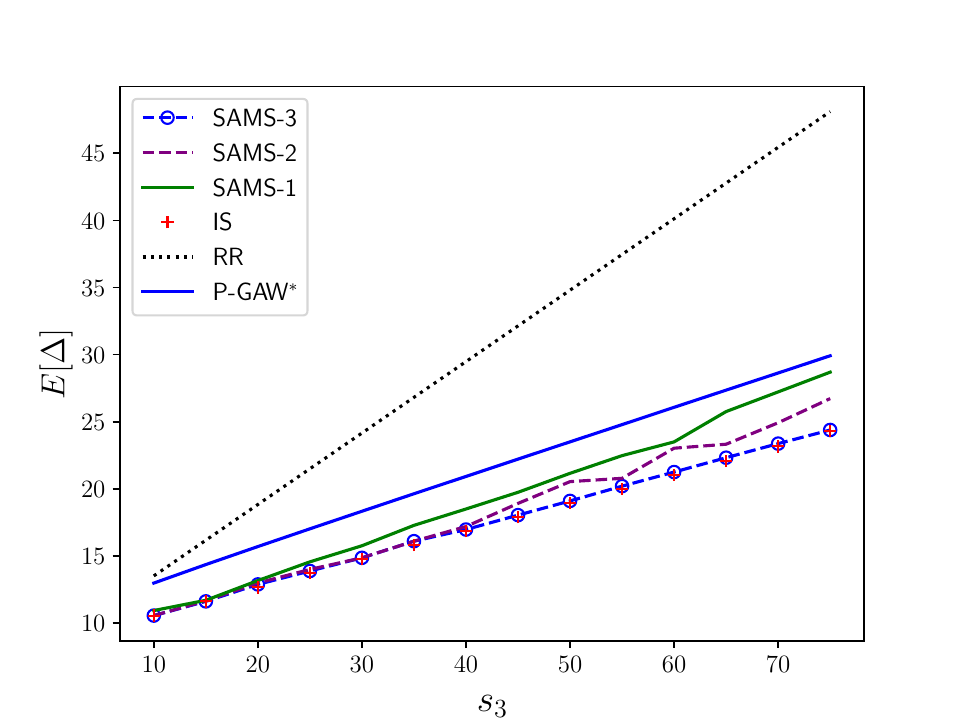}
     \vspace*{-0.2cm}
    \caption{System AoI $\mathbb{E} [\Delta]$ as a function of $s_3$ obtained under various policies.}
    \label{fig:Example1}
\end{figure}
    
\begin{figure}[t]
    \centering
     \vspace*{-0.5cm}
    \includegraphics[width=0.42\textwidth]{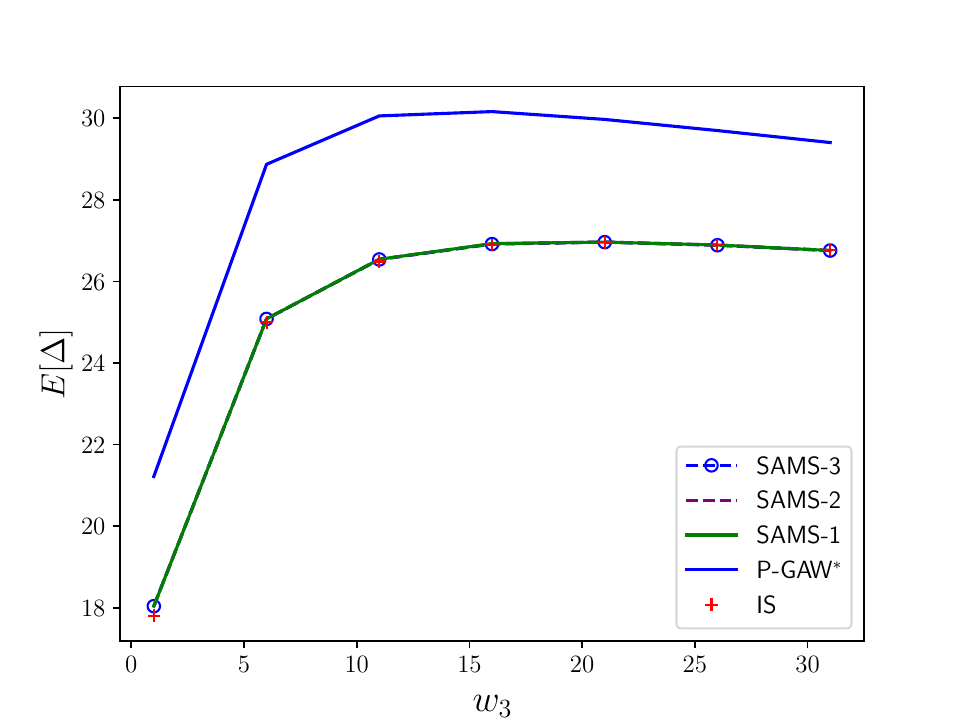}
     \vspace*{-0.2cm}
    \caption{System AoI $\mathbb{E} [\Delta]$ as a function of $w_3$ obtained under various policies.}
    \label{fig:Example2}
\end{figure}

The system AoI as a function of $s_3$ is depicted in Fig.~\ref{fig:Example1} for all the studied policies for this highly heterogeneous scenario for which RR performs very poorly especially when heterogeneity increases, i.e., $s_3$ is increased. 
In this exmaple, we take $I$ to be equal to the size of the pattern produced by SAMS-3. 
The performance gap between P-GAW$^*$ and IS is substantial as also demonstrated in \cite{gamgam_etal_arxiv23}. The results obtained with the three variations of SAMS reveal that (i) exhaustive search over a number of $\varepsilon$ values is advantageous compared to using a single value of this parameter (observe that SAMS-2 outperforms SAMS-1) (ii) with the addition of a few fixed-point iterations, the system AoI performance can further be improved (observe that SAMS-3 outperforms SAMS-2). Overall, for this example, the performance of SAMS-3 has been very close to that of IS.  

In the second numerical example, we evaluate the performance of SAMS in the presence of packet errors, for $N=3$ and deterministic service times given by $s_1 = 10$, $s_2 = s_3=1$. In this experiment, we fix the weights of the first two sources at $w_1 = 2$, $w_2 = 5$ and vary the weight $w_3$ of the third source. 
In this example, we fixed the parameter $I$ to 75 for the IS algorithm. Fig.~\ref{fig:Example2} depicts the system AoI as function of $w_3$ when error probabilities of the 3 sources are given by $p_1= 0.1$, $p_2=0.5$ and $p_3 =0.95$. The RR policy was purposefully avoided in this example, since it is inherently worse compared to all the other policies. In this example, all three variations of the SAMS algorithm presented very similar performance and they also closely followed the IS algorithm while exhibiting a significant performance gain over P-GAW$^*$. Note that the computational complexity of IS is $O(N^2 {I}^3)$ which is far higher than the worst case complexity of $O(N I)$ for SAMS (assuming the produced pattern size $K$ is less than $I$), the latter being applicable to very large-scale scenarios as well. In this regard, SAMS-3 performance comes very close to IS, despite the substantial gap in their computational complexities. 

\begin{figure}[t]
    \centering
    \vspace*{-0.5cm}
    \includegraphics[width=0.41\textwidth]{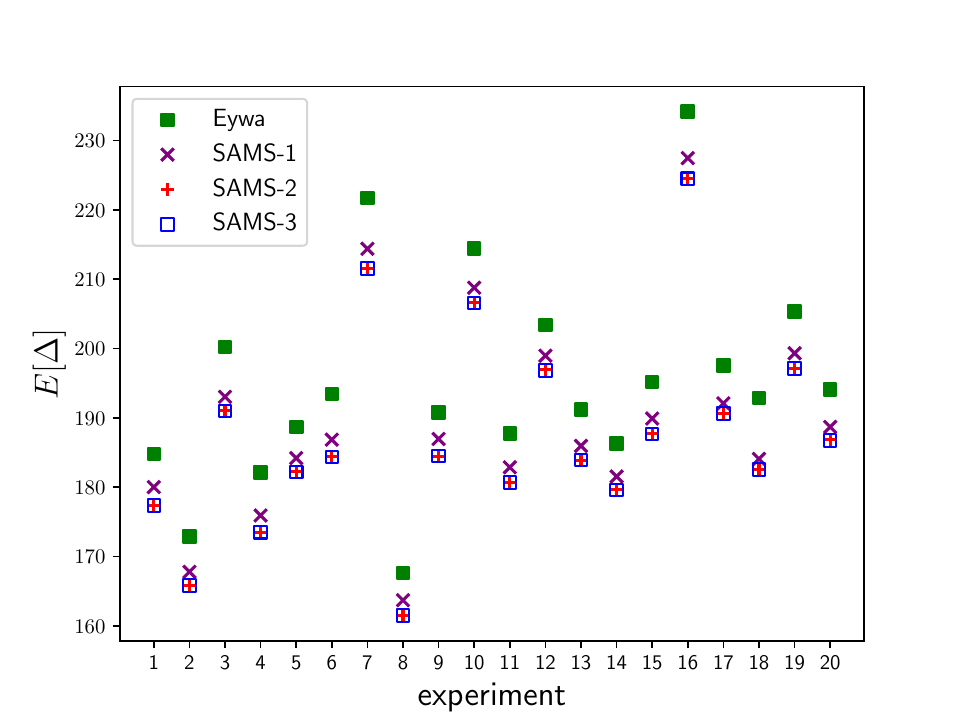}
    \caption{System AoI $\mathbb{E} [\Delta]$ for $N=100$ with deterministic and identical service times for 20 different experiments.}
    \label{fig:Eywa}
\end{figure}

Next, to evaluate the performance of the SAMS algorithm for large scale status update systems, we compared the performance of SAMS against the \emph{Eywa} framework introduced in \cite{eywa} which is designed only for systems with deterministic and identical service times. Therefore, to bring our model to the same domain as \emph{Eywa}, we set the service times of all sources to one with $c_n=0, \ \forall n$. We take $N=100$ sources, and randomly sample 20 weight vectors and probability of error vectors, and evaluate the system AoI of the two schemes, namely SAMS (with its three variations) and \emph{Eywa}, in each case. As illustrated in Fig.~\ref{fig:Eywa}, in each of the experiments, SAMS outperforms \emph{Eywa} by a significant margin. This performance gain is achieved without any sacrifice in the computational complexity. In fact, the computational complexity of \emph{Eywa} is $O(N^2I^2)$. Moreover, in this example, SAMS-2 and SAMS-3 performances are very close. Actually, the $\tilde{c}_n$ values obtained by the SAMS-2 scheduler were already very close to $p_n$ and therefore, there was not much need for performing fixed-point iterations specific to SAMS-3. Finally, the computational complexity of Eywa is much higher than SAMS and this was the sole reason for limiting this particular example only to 20 experiments and 100 sources. Actually, it is quite possible to obtain well-performing schedulers for thousands of information sources with SAMS, thanks to its low computational complexity. 

\begin{figure}[t]
\vspace*{-0.5cm}
    \centering
    \includegraphics[width=0.41\textwidth]{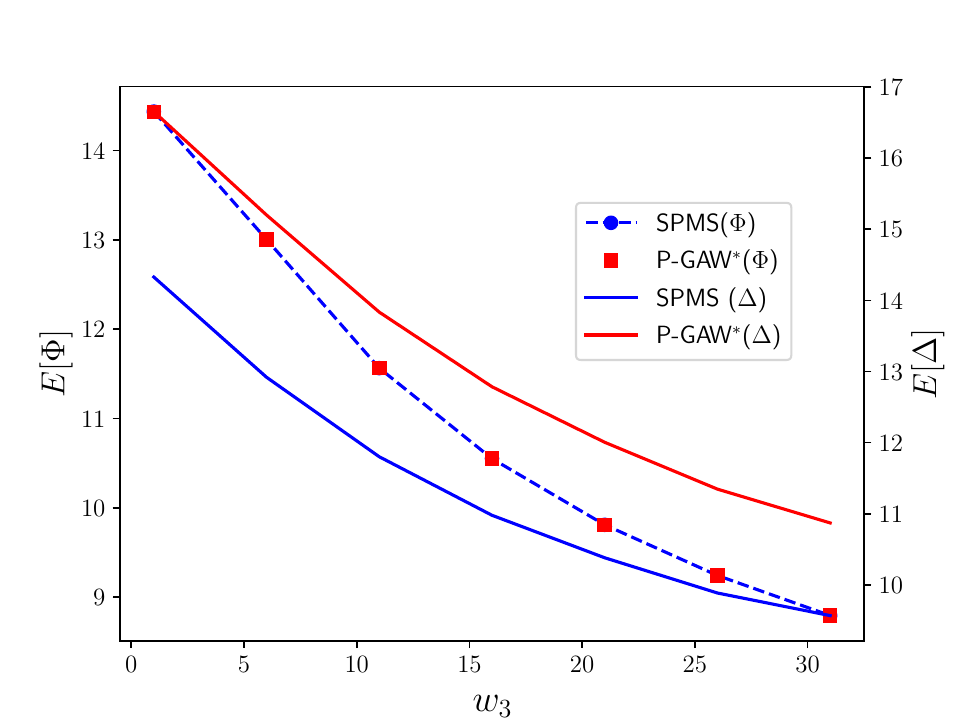}
    \vspace*{-0.3cm}
    \caption{Variation of system PAoI $\mathbb{E} [\Phi]$ and system AoI $\mathbb{E} [\Delta]$ as a function of $w_3$ with $N=3$ and exponential service times.}
    \label{fig:paoi}
\end{figure}

Finally, we compare the performance of the proposed SPMS scheduler against the P-GAW$^*$ scheduler employing $r_n = f_n^*$ derived in \eqref{eqn:optimum_frequency} for the minimization of system PAoI for $N=3$. In this experiment, we use exponentially distributed service times with $[s_1,s_2,s_3] = [10,1,1]$ and error probabilities set to $[p_1,p_2,p_3]=[0.1,0.5,0.6]$. Here, we fix the weights $w_1=2$, $w_2=5$ and vary $w_3$. For each $w_3$, $\varepsilon=2$ is chosen when employing  Algorithm~\ref{alg:General}. As shown in Fig.~\ref{fig:paoi} which depicts the system PAoI as well as the system AoI, as a function of $w_3$, even though the system PAoI of SPMS is comparable to that of P-GAW$^*$, in terms of system AoI, the performance of SPMS far exceeds that of the P-GAW$^*$ scheduler.

\section{Summary} \label{sec:summary}
We have developed a low complexity framework for the development of cyclic schedulers for large-scale status update systems. 
For deterministic service times and in the setting of a small-scale problem with $N=3$, we have shown that our proposed algorithm SAMS performs very close to the recently proposed IS (Insertion Search) based scheduler in terms of system AoI, with SAMS having substantially higher computational efficiency. Observing that the IS algorithm can only be used for small-scale problems, 
we have shown for $N=100$ (as a representative example for a relatively larger size problem) that our framework can generate cyclic schedules that outperform other existing scheduler design frameworks by a good margin in terms of system AoI with much lower computational load. 
We have also shown the benefits of using cyclic scheduling with the proposed SPMS scheduler which minimizes the system PAoI while reducing the system AoI.



\end{document}